NANO EXPRESS

# Electronic States and Light Absorption in a Cylindrical Quantum Dot Having Thin Falciform Cross Section

**Karen G. Dvoyan · David B. Hayrapetyan · Eduard M. Kazaryan · Ani A. Tshantshapanyan**

**Abstract** Energy level structure and direct light absorption in a cylindrical quantum dot (CQD), having thin falciform cross section, are studied within the framework of the adiabatic approximation. An analytical expression for the energy spectrum of the particle is obtained. For the one-dimensional "fast" subsystem, an oscillatory dependence of the wave function amplitude on the cross section parameters is revealed. For treatment of the "slow" subsystem, parabolic and modified Pöschl-Teller effective potentials are used. It is shown that the low-energy levels of the spectrum are equidistant. In the strong quantization regime, the absorption coefficient and edge frequencies are calculated. Selection rules for the corresponding quantum transitions are obtained.

## Introduction

Optical experiments with self-assembled quantum dots (QDs) have demonstrated strong carrier confinement. This is due to the fact that the dot size reduction results in strong "blue shift" of extremely narrow luminescence peaks of isolated dots [1–3]. Confinement effects in magneto-capacitance and infrared absorption have also been observed experimentally [4, 5].

Physical properties of so-called "quantum lenses," or lenticular QDs are of special interest [4, 6, 7]. In particular, energy spectrum of charge carriers (CCs) inside QDs shaped as a spherical segment or an ellipsoid is studied. In reference [8], a cylindrical quantum lens with almost semi-circle cross section was considered. Up to date, however, cylindrical quantum dots (CQDs) with thin lenticular cross sections were studied in paper [9] only.

Typically, a lens geometry is assumed [10], with a circular cross section of maximum radius $r$, and maximum thickness $h$, wherein the CCs are confined in a hard wall potential. Mathematical description of energy levels of such nanostructures is a delicate problem, particularly in the thin lens limit, $h/r \to 0$, which corresponds to a singular perturbation regime.

Study of CQDs having thin falciform cross section will enable one to model more realistic structures which are usually formed in the course of manufacturing. Generally, during growth of QDs, due to unavoidable diffusion process of interface atoms, a coating interlayer between the CQD material and semiconductor matrix is formed. This new interlayer, CDQ with thin falciform cross section, affects the distribution of quantum levels of the CQD significantly.

In this paper, we study electronic states and direct light absorption in CDQs having thin falciform cross section. For the lower energy levels of the CQD, the confining potential is approximated by one-dimensional potential with variable width.

## Theory

Thus, we consider an impenetrable CQD having thin falciform cross section, as shown in Fig. 1a. Potential energy

K. G. Dvoyan · D. B. Hayrapetyan · E. M. Kazaryan · A. A. Tshantshapanyan (✉)
Department of Applied Physics and Engineering, Russian-Armenian State University, 123 Hovsep Emin Str., Yerevan 0051, Armenia
e-mail: achanch@gmail.com

D. B. Hayrapetyan
Department of Physics, State Engineering University of Armenia, 105 Terian Str., Yerevan 0009, Armenia

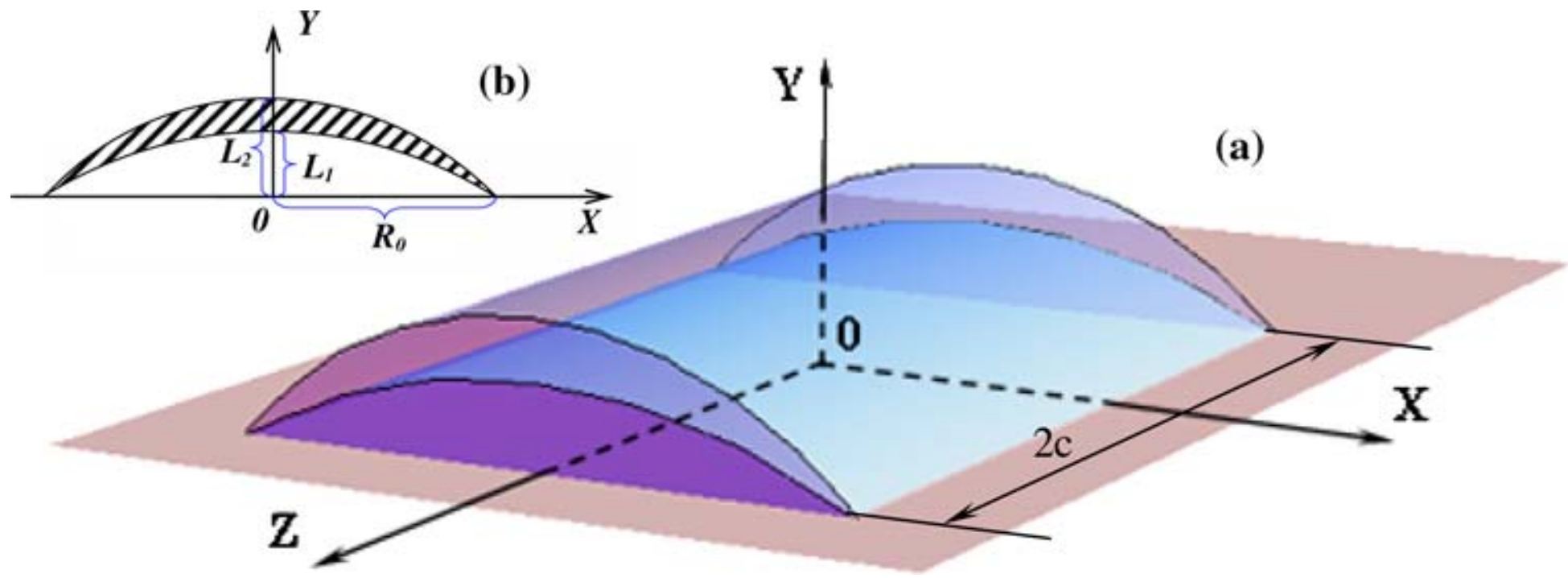

**Fig. 1** **a** Cylindrical quantum dot with thin falciform cross section. **b** Cross section of cylindrical quantum dot

of a charged particle (an electron, or a hole) has the following form:

$$U(x,y,z)=\begin{cases}0; & x^2+(y+y_1)^2\geq R_1^2\cap x^2 \\ & +(y+y_2)^2\leq R_2^2, \\ & z\in[-c,c], \\ \infty; & \text{in the other areas}\end{cases}, \quad (1)$$

where $R_1=(R_0^2+L_1^2)/2L_1$, $R_2=(R_0^2+L_2^2)/2L_2$ are radii of two circles of the cross section, respectively, $2c$ is height of the cylinder, $L_1$, $L_2$ are heights of cross section segments, respectively, $R_0$ is the intersection point of the circles and $X$-axes. The motion of the charged particle in the plane of cross section is localized in the dashed area as it is shown in Fig. 1b.

In the strong size quantization (SQ) regime, the electron-hole Coulomb interaction energy is much less than the confinement energy of the CQD walls. In this regime, one can neglect the Coulomb interaction. Thus, the energy states of the electron and the hole should be determined independently. The particular shape of CQD suggests that motion of a CC in the $Y$-direction should be faster than that in $X$-direction what enables one to apply adiabatic approximation. The Hamiltonian of the system in this case has the following form:

$$\hat{H}=-\frac{\hbar^2}{2\mu}\left[\frac{\partial^2}{\partial X^2}+\frac{\partial^2}{\partial Y^2}+\frac{\partial^2}{\partial Z^2}\right]+U(X,Y,Z). \quad (2)$$

Being expressed through dimensionless variables, the Hamiltonian (2) may be represented as the sum of the "fast" and "slow" subsystems' operators, $\hat{H}_1$ and $\hat{H}_2$, respectively, and, the $Z$-direction $\hat{H}_3$ operator:

$$\hat{H}=\hat{H}_1+\hat{H}_2+\hat{H}_3+U(x,y,z), \quad (3)$$

where

$$\hat{H}_1=-\frac{\partial^2}{\partial y^2},\ \hat{H}_2=-\frac{\partial^2}{\partial x^2},\ \hat{H}_3=-\frac{\partial^2}{\partial z^2}, \quad (4)$$

and the following notations are introduced: $x=X/a_B$, $y=Y/a_B$, $z=Z/a_B$, $\hat{H}=\hat{H}/E_R$, with $E_R=\hbar^2/2m_e a_B^2$ being the effective Rydberg energy, $a_B=\kappa\hbar^2/m_e e^2$, the effective Bohr radius of electron, $m_e$, the effective mass of electron, and $\kappa$, the dielectric constant of the medium. We seek the wave function of the problem in the following form:

$$\Psi(x,y,z)=\varphi(x)f(y;x)\chi(z). \quad (5)$$

Due to the CQD problem symmetry, motion of the electron in the $z$-direction is separated. The energy is given by the following expression:

$$\varepsilon_z=\frac{\pi^2 n_z^2}{4c^2},\quad n_z=1,2,\ldots, \quad (6)$$

where $n_z$ is the quantum number.

When the coordinate of the "slow" subsystem, $x$, is fixed, the motion of the electron is localized in the one-dimensional effective potential well, having the following spatial profile:

$$h(x)=h_2(x)-h_1(x), \quad (7)$$

$$\begin{aligned} h_1(x)&=\sqrt{\frac{\left(R_0^2+L_1^2\right)^2}{4L_1^2}-x^2}+L_1-\frac{R_0^2+L_1^2}{2L_1}, \\ h_2(x)&=\sqrt{\frac{\left(R_0^2+L_2^2\right)^2}{4L_2^2}-x^2}+L_2-\frac{R_0^2+L_2^2}{2L_2}, \end{aligned} \quad (8)$$

where $L\equiv L_2-L_1$ is the maximal value of CQD falciform cross section height.

The Schrödinger equation for the "fast" subsystem has the form

$$f(y;x)''+f(y;x)\varepsilon_1(x)=0. \quad (9)$$

After simple transformations, one can obtain the following expressions for the wave function and electron energy, respectively:

$$\begin{aligned} f(y;x) &= \sqrt{\frac{2}{h_2(x)-h_1(x)}}\cos\left(\frac{\pi n h_1(x)}{h_1(x)-h_2(x)}\right)\sin\frac{\pi n}{h_1(x)-h_2(x)}y \\ &+\sqrt{\frac{2}{h_2(x)-h_1(x)}}\sin\left(\frac{\pi n h_1(x)}{h_1(x)-h_2(x)}\right)\cos\frac{\pi n}{h_1(x)-h_2(x)}y, \end{aligned} \quad (10)$$

$$\varepsilon_1(x) = V_{\text{Real}} = \frac{\pi^2 n^2}{h^2(x)}, \quad n = 1, 2, \ldots, \tag{11}$$

where $n$ is the quantum number.

Here we obtain the following result: the wave functions' amplitudes depend on geometrical parameters of the QD shape. This means that probability of the CC localization presents oscillatory behavior near the peripheral areas of CQD.

Expression (11) takes the place of the potential in the Schrödinger equation for the "slow" subsystem, but the Schrödinger equation with such effective potential is not analytically solvable. That is why we have applied adiabatic approximation, to solve this problem. Two models for the "slow" subsystem effective potentials are used.

### Parabolic Approximation

The "slow" subsystem potential energy is formed by falciform geometry of the QD cross section which allows to use adiabatic approximation. Namely, we use the condition $|x| \ll R_0$ which means that CC is localized in the vicinity of the geometric center of falciform cross section. This condition holds for the low-energy states. For the higher excited energy states, the adiabatic approximation is not applicable. The energy of the "fast" subsystem is represented by the Taylor series, where linear, cubic, and other odd terms are equal to zero:

$$\varepsilon_1(x) = V_{\text{Par}}(x) \approx \alpha_n + \beta_n^2 x^2 \tag{12}$$

with

$$\alpha_n = \frac{\pi^2 n^2}{L^2}, \quad \beta_n^2 = \frac{\pi^2 n^2 (R_1 - R_2)}{R_1 R_2 L^3}. \tag{13}$$

Condition $L \ll R_0$ clearly indicates that the fourth-order term is about 100 times smaller than the quadratic term. Here it should be noted that the origin of the quadratic potential is due to the fact that the width of one-dimensional effective potential well is a variable quantity.

Expression (12) plays the role of the effective potential in the Schrödinger equation for the "slow" subsystem:

$$\varphi''(x) + \left(\varepsilon - \alpha_n - \beta_n^2 x^2\right)\varphi(x) = 0. \tag{14}$$

Solving this equation, we obtain the expressions for the CC wave function and energy:

$$\varphi(x) = e^{-\frac{\beta_n x^2}{2}} H_N\left(\sqrt{\beta_n} x\right), \tag{15}$$

$$\varepsilon = \alpha_n + 2\beta_n\left(N + \frac{1}{2}\right), \quad N = 0, 1, 2, \ldots \tag{16}$$

where $H_N\left(\sqrt{\beta_n} x\right)$ are Hermit polynomials, and $N$ is oscillatory quantum number.

### Modified Pöschl-Teller Potential Approximation

As it was mentioned above, the adiabatic approximation is applicable for calculation of lower levels of the energy spectrum. Parabolic potential, obtained by use of Taylor series of the energy expression for the "fast" subsystem, gives rise to a set of equidistant energy levels in spectrum. It is notable that each energy level of the "fast" subsystem has its own set of equidistant levels with gaps depending on the quantum number of the particular "fast" subsystem. However, only two or three lower energy levels are split into equidistant level subsystems; for higher levels of the "fast" subsystem the sublevels are not equidistant any more.

We suggest a more realistic model of one-dimensional effective potential which we represent in the form of modified Pöschl-Teller potential (see Fig. 2) [11, 12]. In dimensionless quantities, this potential has the following form:

$$\varepsilon_1(x) = V_{\text{PT}}(x) = \frac{\pi^2 n^2}{L^2} - \frac{\lambda(\lambda - 1)}{\gamma^2 (ch(x/\gamma))^2} + \frac{\lambda(\lambda - 1)}{\gamma^2}. \tag{17}$$

Here $\lambda$ and $\gamma$ are parameters describing the depth and width of corresponding quantum well, respectively. Note that they depend on the quantum number $n$ of the "fast" subsystem. Choice of this particular potential is explained by the fact that the Taylor expansion of potential (17) for small values of the $x$-coordinate is parabolic as it is the case for (12) also. On the other hand, at higher values of the $x$-coordinate the discrepancy of the Pöschl-Teller potential from parabolic one is increasing. Thus, violation of equidistance of energy levels of "slow" subsystem can be taken into account.

The one-dimensional Schrödinger equation with the Pöschl-Teller potential reads:

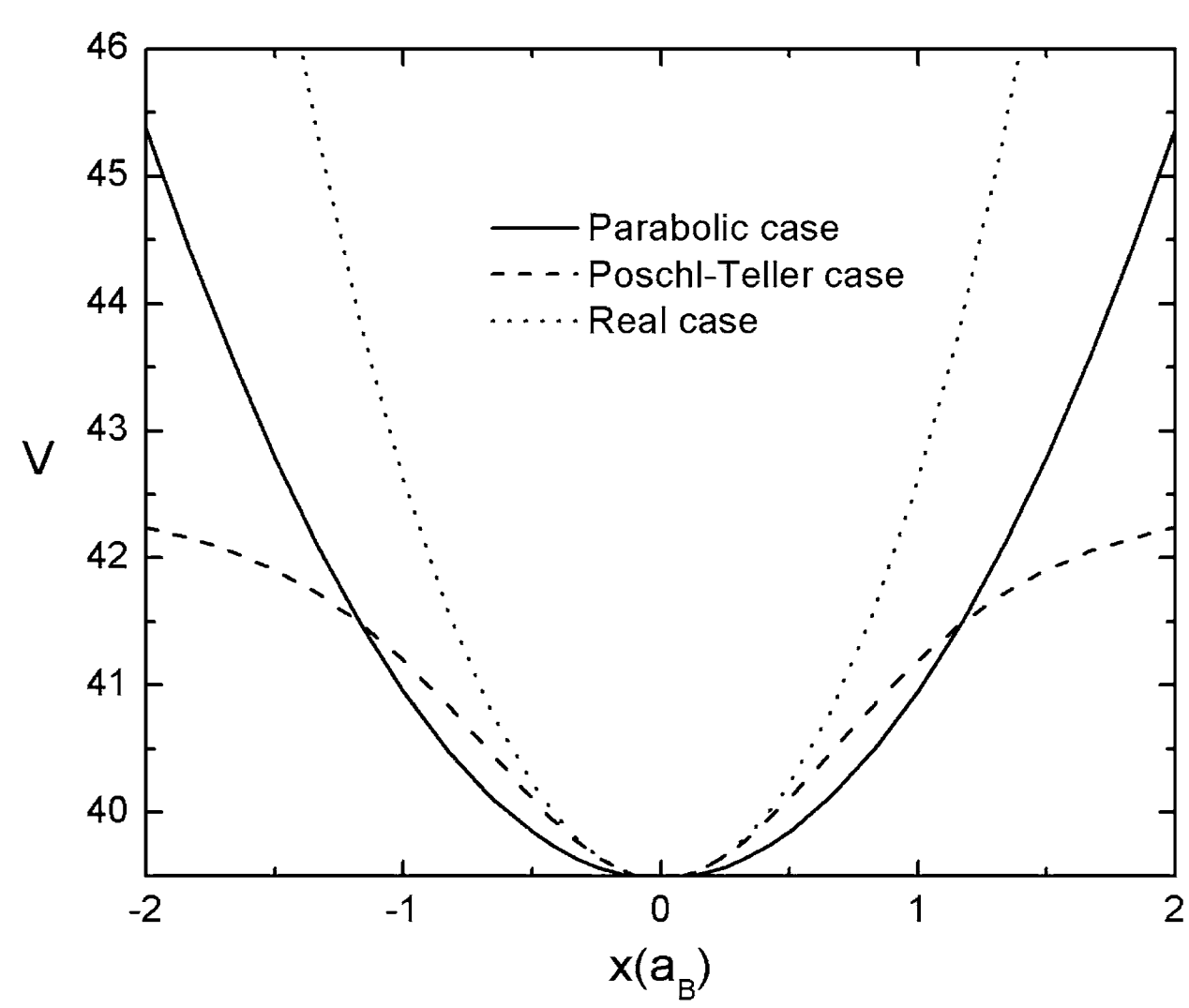

**Fig. 2** Dependence of one-dimensional effective potentials on coordinate $x$

$$\varphi''(x)+\left(\varepsilon-\frac{\pi^2n^2}{(L_2-L_1)^2}-\frac{\lambda(\lambda-1)}{\gamma^2}+\frac{\lambda(\lambda-1)}{\gamma^2(ch(x/\gamma))^2}\right)\varphi(x)=0. \tag{18}$$

We adopt the following notation:

$$k^2=\varepsilon-\frac{\pi^2n^2}{L^2}-\frac{\lambda(\lambda-1)}{\gamma^2} \tag{19}$$

thus reducing Eq. 18 to the following one [12]:

$$\varphi''+\left(k^2+\frac{\lambda(\lambda-1)}{\gamma^2(ch(x/\gamma))^2}\right)\varphi=0. \tag{20}$$

A series of transformations results in the following expressions for the wave function and energy spectrum of CC:

$$\begin{aligned}\varphi(x)=ch^{\lambda}\left(\frac{x}{\gamma}\right)\Bigg[&C_1\cdot{}_2F_1\left(u,v,\frac{1}{2};1-ch^2\left(\frac{x}{\gamma}\right)\right)\\&+C_2\left(1-ch^2\left(\frac{x}{\gamma}\right)\right)^{\frac{1}{2}}\\&\cdot{}_2F_1\left(u+\frac{1}{2},v+\frac{1}{2},\frac{3}{2};1-ch^2\left(\frac{x}{\gamma}\right)\right)\Bigg],\end{aligned} \tag{21}$$

$$\varepsilon=\frac{(\lambda-1-n)^2}{\gamma^2}+\frac{\pi^2n^2}{L^2}+\frac{\lambda(\lambda-1)}{\gamma^2}, \tag{22}$$

where $u=\frac{1}{2}(\lambda-\gamma k)$ and $v=\frac{1}{2}(\lambda+\gamma k)$, $C_1$ and $C_2$ are normalization constants, ${}_2F_1(a,b,c;x)$ is the hypergeometric function. For small values of the coordinate $x$, the potential (17) takes the form

$$V_{\mathrm{PT}}(x)\approx\frac{\pi^2n^2}{L^2}+\frac{\lambda(\lambda-1)x^2}{\gamma^4}. \tag{23}$$

Further, solution to the Schrödinger equation for the "slow" subsystem with the potential (23) is completely similar to the procedure with parabolic potential considered above. As a result, we arrive at the following expression for the equidistant energy spectrum of a CC:

$$\varepsilon=\alpha_n+\frac{2\sqrt{\lambda(\lambda-1)}}{\gamma^2}\left(N+\frac{1}{2}\right),\quad N=0,1,2,\ldots \tag{24}$$

which perfectly agrees with the result (16).

## Direct Interband Light Absorption

Now, we consider direct interband light absorption by CQD with thin falciform cross section, in the strong SQ regime. This means that the conditions $L\ll\left\{a_B^e,a_B^h\right\}$ are satisfied, where $a_B^{e(h)}$ is an effective Bohr radius of the electron (or the hole). We consider the case of a heavy hole, when $m_e\ll m_h$, with $m_e$ and $m_h$ being the effective masses of the electron and hole, respectively. Under conditions of one-electron band theory approximation, the absorption coefficient is given by the expression [13]:

$$K=A\sum_{\nu,\nu'}\left|\int\Psi_\nu^e\Psi_{\nu'}^h\,d\vec{r}\right|^2\delta\left(\hbar\Omega-E_g-E_\nu^e-E_{\nu'}^h\right), \tag{25}$$

where $\nu$ and $\nu'$ are sets of quantum numbers corresponding to the electron and heavy hole, $E_g$ is the forbidden gap width in the bulk semiconductor, $\Omega$ is the incident light frequency, and $A$ is a quantity proportional to the square of matrix element in decomposition over Bloch functions.

We have performed numerical simulations for a QD consisting of *GaAs* with the following parameters: $m_e=0.067m_0$, $m_e=0,12m_h$, $E_R=5.275\,\mathrm{meV}$, $a_B^e=104$ Å, $E_g=1.43\,\mathrm{eV}$, $m_0$ is a free electron mass. Finally, for the parabolic case for the quantity $K$ and, absorption edge (AE) we obtain, respectively,

$$K=A\sum_{n,N,N'}I_{n,N,N'}\delta\left(\hbar\Omega-E_g-E_\nu^e-E_{\nu'}^h\right) \tag{26}$$

$$W_{110}=1+\frac{\pi^2}{4}\frac{d^2}{c^2}+\pi^2\frac{d^2}{L^2}+2\pi\sqrt{\frac{R_2-R_1}{R_1R_2}}\frac{d^2}{\sqrt{L^3}} \tag{27}$$

where $W_{110}=\hbar\Omega_{110}/E_g$, $I_{n,N,N'}$ is an integral quantity (see Appendix 1). Here we use expression $d=\hbar/\sqrt{2\mu E_g}$ as a length unit, where $\mu=m_em_h/m_e+m_h$ is the reduced mass of the electron and hole.

Selection rules in the case of the parabolic potential appear to be as follows: $n_z=n_z'$. For other quantum numbers transitions between the corresponding levels are admissible. Note that in the limit case when $L_1\to0$ the falciform cross section becomes a segment of a circle and we arrive at the following well-known result: the transitions are allowed between the energy levels having quantum numbers in $z$-direction $n_z=n_z'$, in $y$-direction, $n=n'$ and, different parity (see [8]). In the oscillatory quantum number values, transitions are allowed between the levels either having $N=N'$ and equal parity quantum numbers, $N-N'=2t$. Partial reduction of number of selection rules in the case of falciform cross section of cylindrical-well QD is due to oscillatory character of the dependence of wave function's amplitude (10) on cross section parameters of the QD. Obviously, that transition to the limit $L_1\to0$ is equivalent to the limit $h_1(x)\to0$ in expression (10). Thus, the oscillatory character of dependence of the wave function amplitude [see (10)] on cross section parameters of QD is canceled. In other words, the electron and hole wave functions' overlap integral in the falciform cross section plane is always nonzero, a fact which partially reduces the selection rules' number.

It is worth mentioning that in the case of cylindrical QD with circular cross section considered in the paper [14], the transitions are allowed between the energy levels with quantum numbers $m = -m'$, $n_z = n'_z$, where $m$ is magnetical quantum number in the plane of cross section.

In the case of modified Pöschl-Teller potential, for the quantity $K$ and AE we obtain

$$K = A \sum_{n,N,N'} J_{n,N,N'} \delta\left(\hbar\Omega - E_g - E_\nu^e - E_{\nu'}^h\right) \tag{28}$$

$$W_{110} = \frac{\pi^2}{4}\frac{d^2}{c^2} + \pi^2 \frac{d^2}{L^2} + (\lambda-2)^2 \frac{d^2}{\gamma^2} + \lambda(\lambda-1)\frac{d^2}{\gamma^2} \tag{29}$$

where $W_{100} = \hbar\Omega_{100}/E_g$, $J_{n,N,N'}$ is an integral quantity (see Appendix 1). In this case the same selection rules for quantum numbers are valid, as in the case of parabolic approximation.

## Discussion of Results

As one can see from the CC energy spectrum expressions (16) and (24), the energy levels inside the CQD with falciform cross section are equidistant. More precisely, each level of the "fast" subsystem has its own family of equidistant energy levels created by the "slow" subsystem. As a consequence of the adiabatic approximation, this result is valid only for the low spectrum levels (i.e., small quantum numbers). Note that the CC levels are equidistant in the case when $h_1(x) \to 0$ also [9]. However, in the case when $h_1(x) \neq 0$, the wave function amplitude dependence on falciform cross section parameters is unique [see (10)]. As it is mentioned above, this dependence results in oscillatory behavior of the wave function amplitude thus affecting the overlap integral form and partially reduces the selection rules set. It is also important that approximation of one-dimensional energy expression by a modified Pöschl-Teller potential enables one to take into account the energy levels which are nonequidistance at higher energy values. One can see from Fig. 2 that the effective one-dimensional potential is well approximated by the modified Pöschl-Teller potential. As the $x$-coordinate grows, the discrepancy between the exact and approximate potentials becomes evident both for the modified Pöschl-Teller and parabolic potentials (see Appendix 2).

Figure 3 illustrates the dependence of the CC spectral energy levels for the first equidistant family inside CQD having falciform cross section as a function of height $L_1$ of the small cross section segment, in both cases of one-dimensional potential approximation. In other words, we compare results obtained from relations (16) with those from (24). From Fig. 3, it is easily seen that the CC energy levels are equidistant in both cases since for small values of the $x$-coordinate it is sufficient to keep only quadratic terms in the Taylor development of the modified Pöschl-Teller potential what leads to practical coincidence with parabolic potential. Growth of the parameter $L_1$ results in width reduction of falciform cross section of the CQD, which in its turn increases the CC energy due to SQ. However, when decomposition of modified Pöschl-Teller potential is used, the energy levels are positioned higher than in parabolic potential approximation and their gap is increased with $L_1$. This fact is explained by higher SQ portion in the particle energy. One can see from Fig. 2, with increasing of $x$-coordinate approximated modified Pöschl-Teller potential increases faster than parabolic potential. Thus the effect of QD walls is stronger in the first case than in the second.

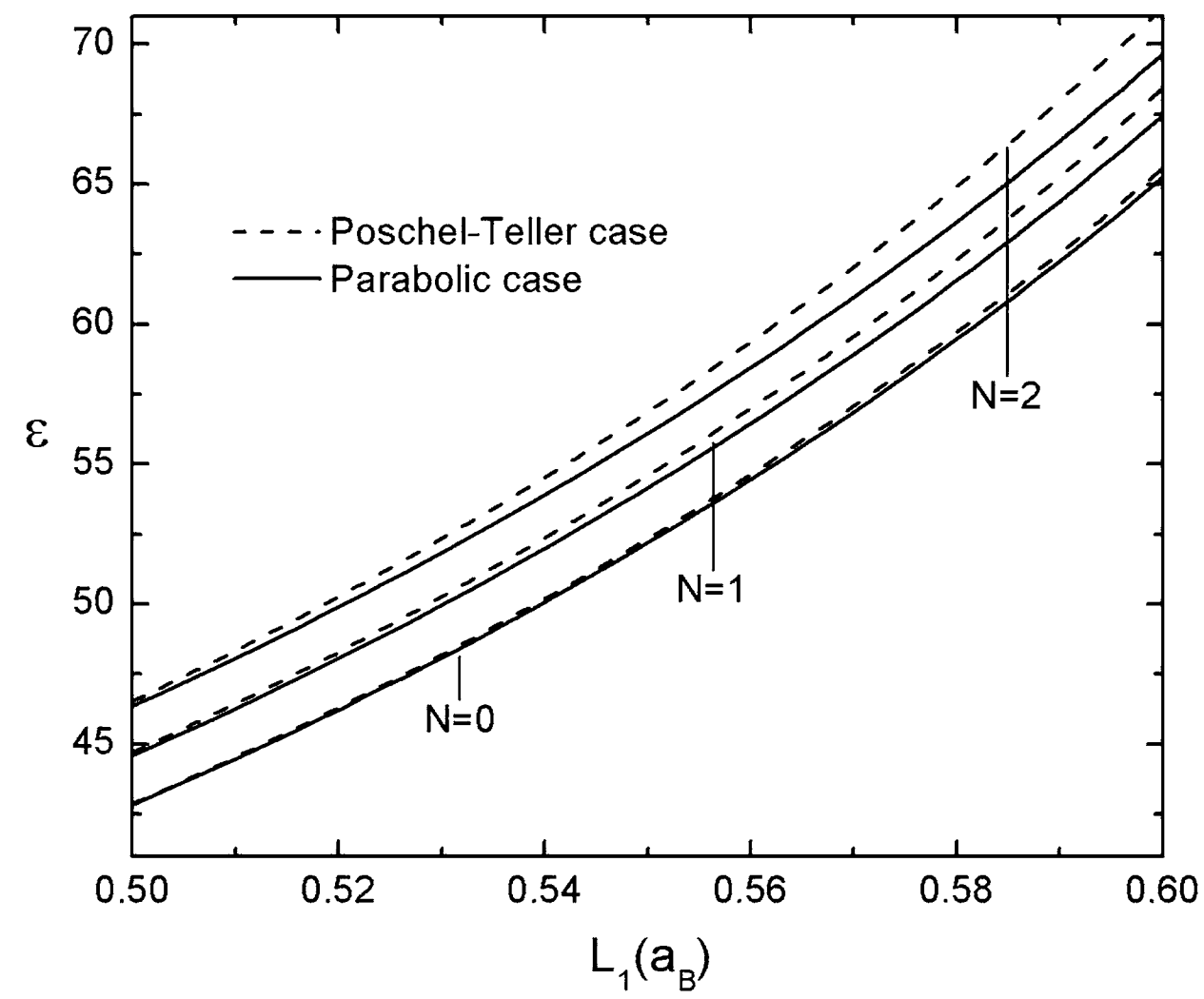

**Fig. 3** Dependence of first equidistant family of CC energy in CQD with thin falciform cross section on height of segment $L_1$ for both parabolic and Pöschl-Teller cases

Figure 4 illustrates the dependence of first three energy levels of a CC on the height $L_1$ of the small segment in the falciform cross section, when the modified Pöschl-Teller potential approximation is used. Note that the energy levels are not more equidistant (see expression 22). As it was mentioned above, the modified Pöschl-Teller potential allows describing nonparabolic character of the CC energy, the fact clearly shown in Fig. 4. Such dependence (both in "fast" and "slow" motions) opens a sufficiently broad opportunity for using the CQD ensemble as an active medium in quantum lasers. For example, in US Patent #6541788 a method and device for converting light from a first wavelength to a second wavelength is presented, where acting objects are multilayer ellipsoidal quantum dots and lenses; it is a good example for targeting applications this research.

Figure 5 shows dependence of light absorption frequency edges for the CQD on the height $L_1$ of the small segment of falciform cross section under fixed values of large segment height $L_2$, when parabolic approximation is used. Note that $L_1$ growth causes the AE shift to higher frequencies ("blue" shift). Thus, the contribution of SQ becomes higher as falciform width is reduced. For the same reason the curves corresponding to small $L_2$ values are positioned higher. Opposite behavior is seen in Fig. 6, where AE dependence on the large segment height $L_2$ under fixed values of small segment height $L_1$ is shown for the case when parabolic approximation is used. As expected, larger $L_2$ values cause the shift of AE to low frequencies ("red" shift). This phenomenon is explained by reduced SQ effect of QD walls when width of the falciform cross section becomes larger. The curves corresponding to small values of parameter $L_1$ are positioned below, which is also explained by reduction of confinement effect.

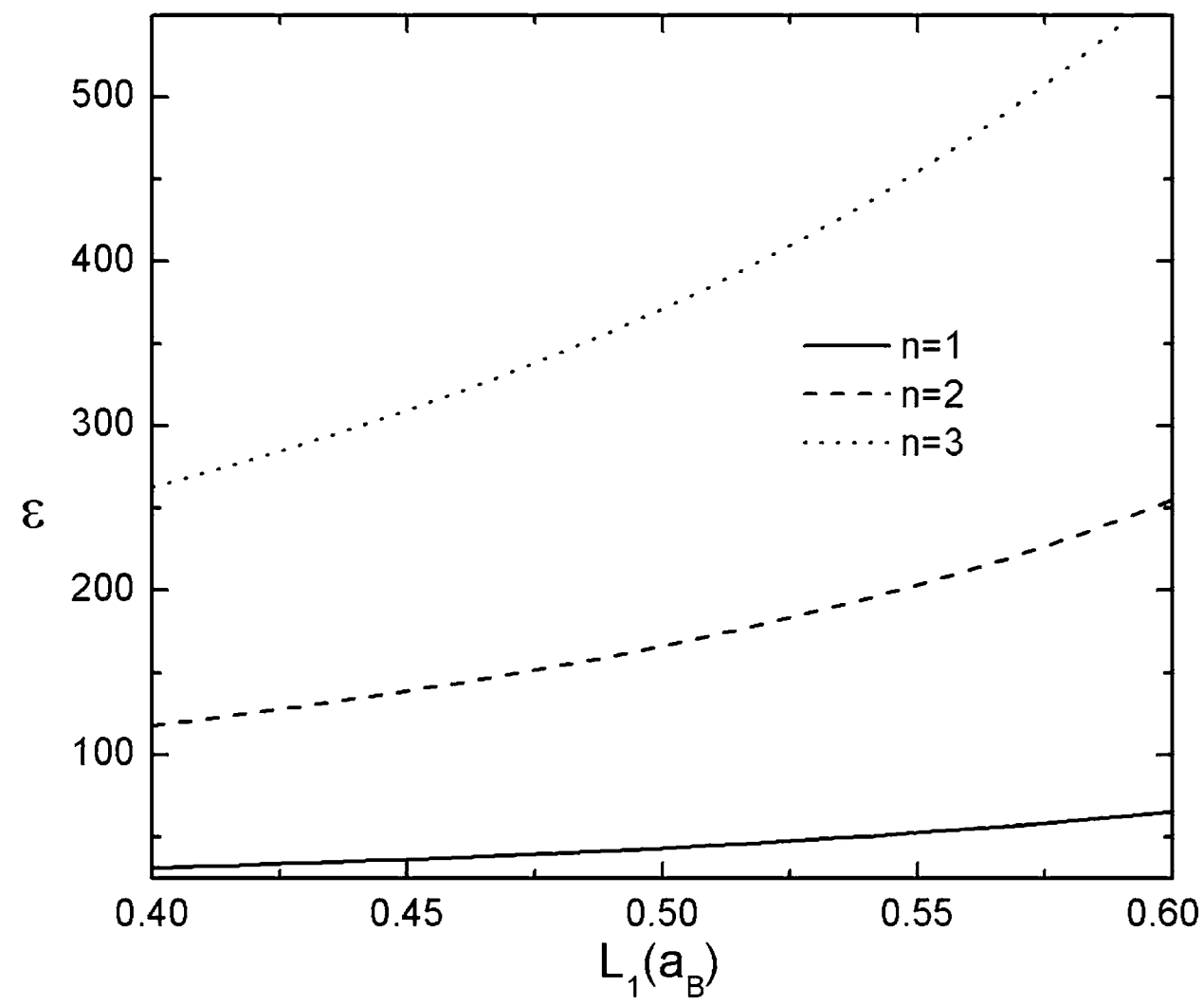

**Fig. 4** Dependence of first three levels of CC energy in CQD with thin falciform cross section on height of segment $L_1$ for Pöschl-Teller potential realization case

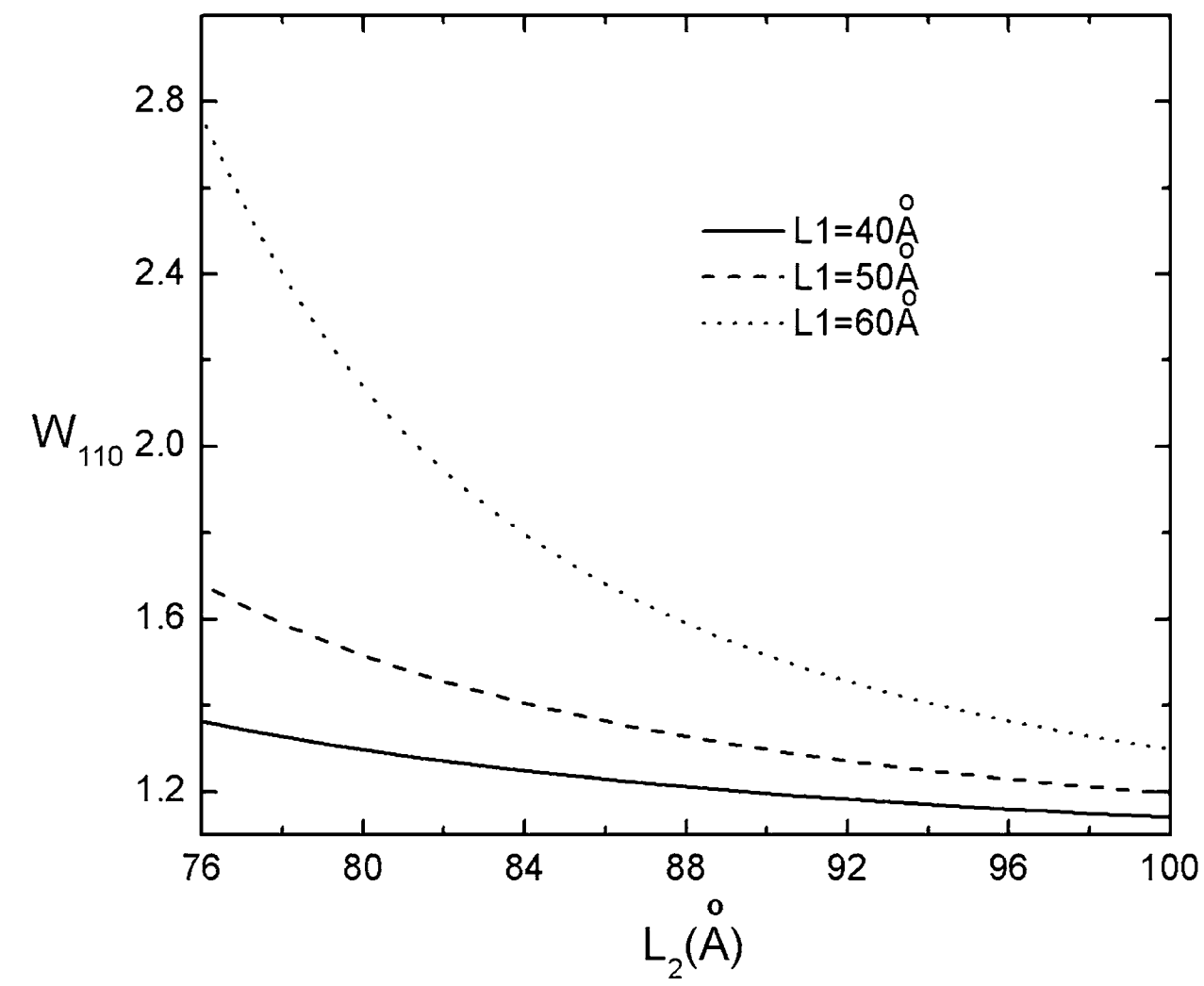

**Fig. 6** Dependence of AE in CQD with thin falciform cross section on height of segment $L_2$ for parabolic potential realization case for fixed values of $L_1$

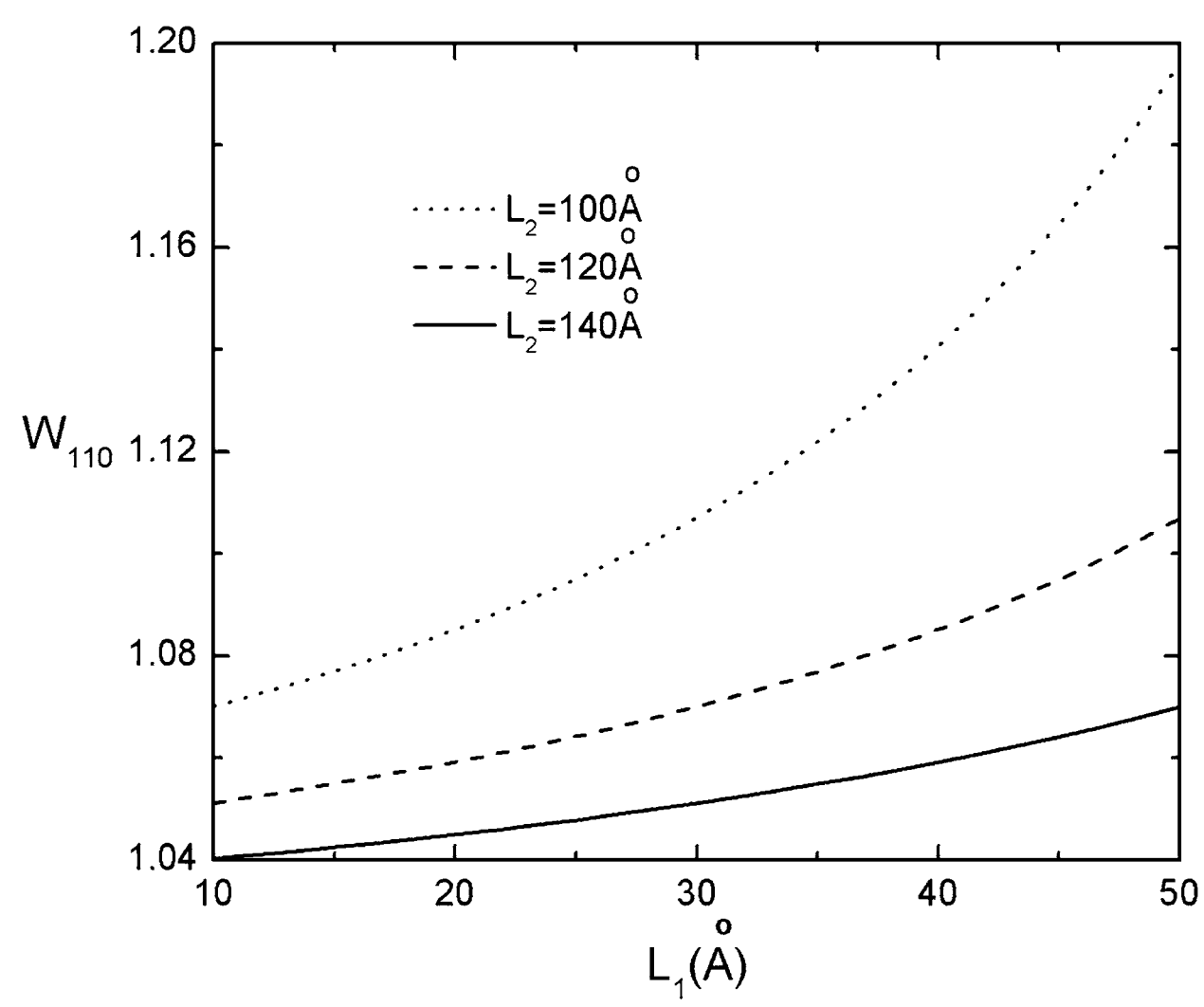

**Fig. 5** Dependence of AE in CQD with thin falciform cross section on height of segment $L_1$ for parabolic potential realization case for fixed values of $L_2$

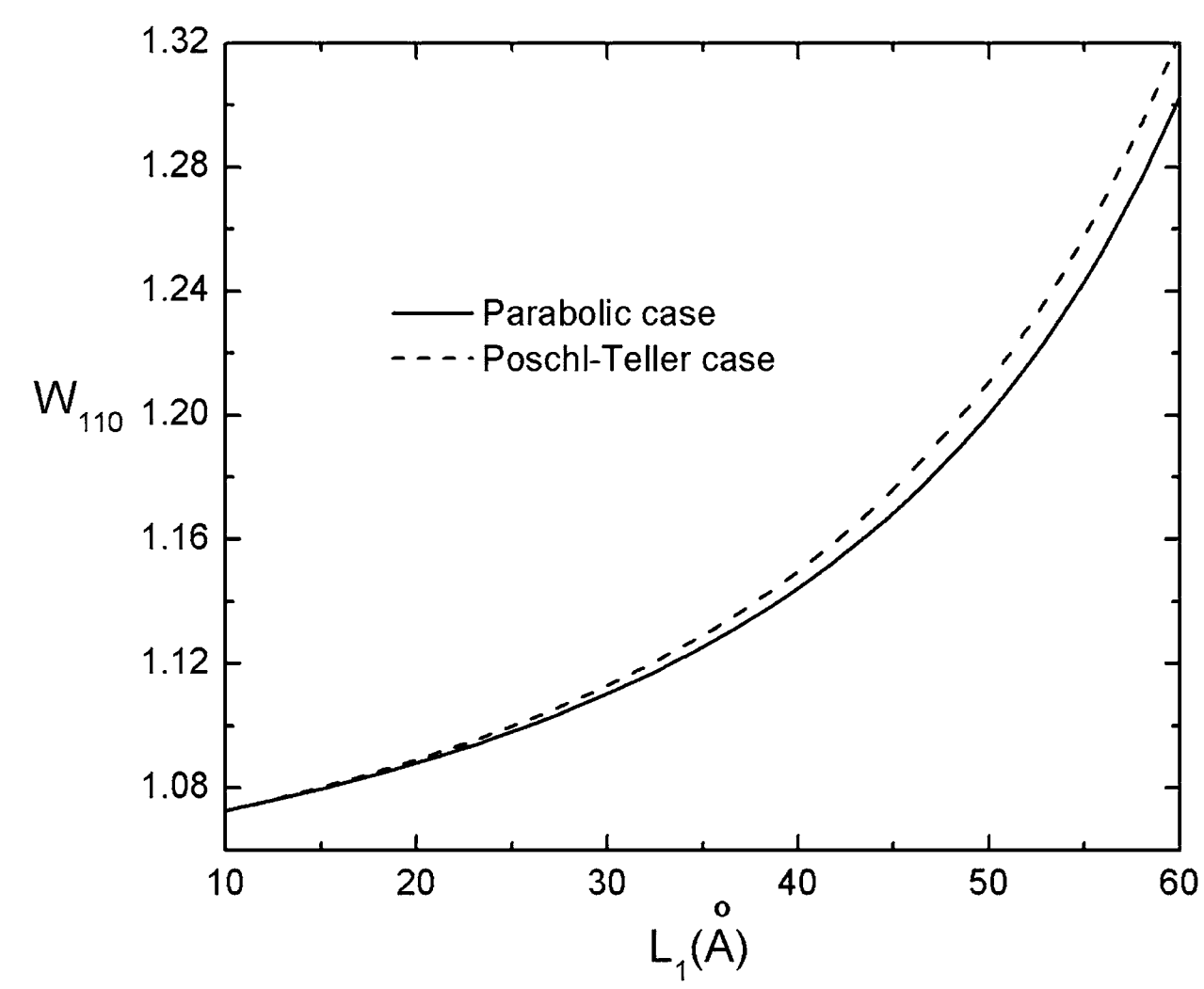

**Fig. 7** Dependences of AE in CQD with thin falciform cross section on height of segment $L_1$ for fixed values of $L_2$ for both parabolic and Pöschl-Teller cases

Finally, in Figs. 7 and 8 comparisons are given of the AE values or the falciform cross section with parameters $L_1$ and $L_2$ in the cases when parabolic and modified Pöschl-Teller potential approximations are used. One can see in the Fig. 7 that the curves converge when $L_1$ is small (broad cross section). As the height $L_1$ is increased, the AE, as it has already been mentioned, shifts to higher frequencies and the difference between the AE values observed more distinctly due to higher contribution of SQ. And vice versa,

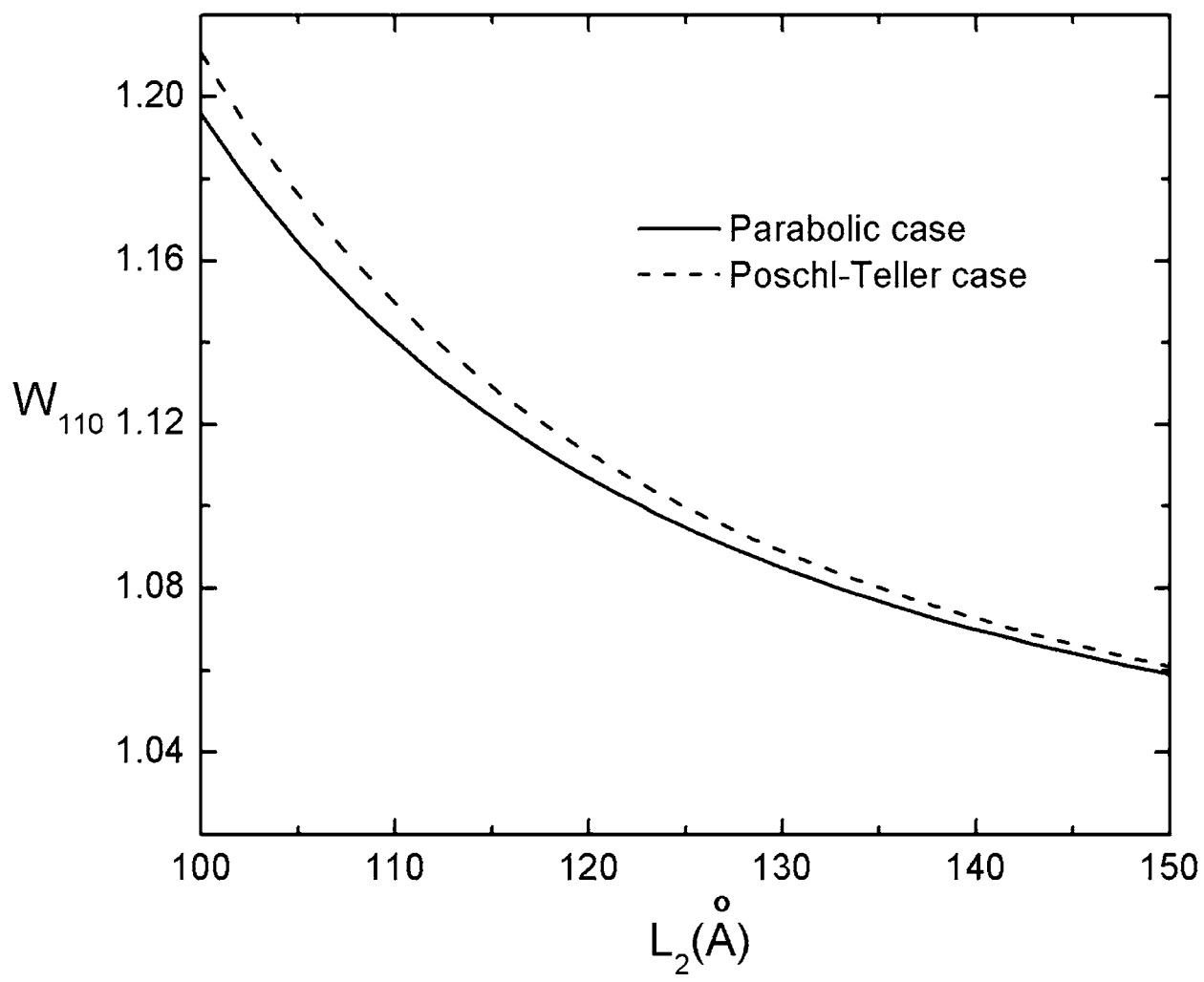

**Fig. 8** Dependences of AE in CQD with thin falciform cross section on height of segment $L_2$ for fixed values of $L_1$ for both parabolic and Pöschl-Teller cases

similar interpretation can be given to Fig. 8 where the AE shifts to low frequencies and curves converge as the parameter $L_2$ is increased.

## Conclusion

Thus, we have theoretically proved that energy spectrum of a CC inside CQD having falciform cross section is equidistant for the lower spectrum levels. Meanwhile, the energy dependence on geometric parameters of QD has the root character. We have revealed the unique (oscillatory) character of the wave function amplitude dependence on geometric parameters of CQD cross section. The formed one-dimensional effective motion potential has been successfully modeled by modified Pöschl-Teller potential, which makes possible an account of the real potential divergence from the parabolic potential. The effect of the former potential which we developed in Taylor series for the lower energy levels of CC (provided that the particle is localized in the cross section center of CQD) has been compared in the paper with the effect of purely parabolic potential. Direct interband light absorption by CQD having falciform cross section has been analyzed. The oscillatory dependence of the effective one-dimensional motion wave function amplitudes on geometric parameters of the cross section has shown lead to partial reduction of selection rules in light absorption process.

Cylindrical quantum lenses and especially falciform CQDs, as a more realistic nanostructures with account of nonparabolicity of forming potential, have various commercial applications. For example, they are widely used in large two-dimensional focal plane arrays in the mid- and far infrared (M&FIR) region. They also have important applications at pollution detection, thermal imaging object location and remote sensing as well as infrared imaging of astronomical objects.

Mentioned optimized quantum structures can be formed by direct epitaxial deposition using a self-assembling QDs technique, described, e.g., in the US Patent #6541788 entitled as “Mid infrared and near infrared light upconverter using self-assembled quantum dots” as well as by usage of MBE, MOCVD, or MOMBE deposition systems.

Results of presented theoretical investigation can be directly applied to the photonics field as background for simulation model. One of the hot topics of this field is developing a scheme for optimization of growth of CQD needed for second harmonic generation.

**Acknowledgments** This research has been undertaken with financial support of ANSEF grant PS-nano #1301 and Armenian State Target Program “Semiconductor Nanoelectronics.”

## Appendix 1

$$I_{n,N,N'} = \left| \int_0^\infty \frac{h_2(x)+h_1(x)-\frac{(h_1(x)-h_2(x))}{2\pi n}\sin\left(\frac{h_2(x)+h_1(x)}{h_1(x)-h_2(x)}2\pi n\right)}{h_2(x)-h_1(x)} \times \exp\left\{-\frac{1}{2}\sqrt{\frac{\pi^2 n^2\left(\frac{R_0^2+L_1^2}{2L_1}-\frac{R_0^2+L_2^2}{2L_2}\right)}{\frac{R_0^2+L_1^2}{2L_1}\frac{R_0^2+L_2^2}{2L_2}(L_2-L_1)^3}}x^2\right\} H_N\left(\sqrt[4]{\frac{\pi^2 n^2\left(\frac{R_0^2+L_1^2}{2L_1}-\frac{R_0^2+L_2^2}{2L_2}\right)}{\frac{R_0^2+L_1^2}{2L_1}\frac{R_0^2+L_2^2}{2L_2}(L_2-L_1)^3}}x\right) \times \exp\left\{-\frac{1}{2}\sqrt{\frac{\pi^2 n^2\left(\frac{R_0^2+L_1^2}{2L_1}-\frac{R_0^2+L_2^2}{2L_2}\right)}{\frac{R_0^2+L_1^2}{2L_1}\frac{R_0^2+L_2^2}{2L_2}(L_2-L_1)^3}}x^2\right\} H_{N'}\left(\sqrt[4]{\frac{\pi^2 n^2\left(\frac{R_0^2+L_1^2}{2L_1}-\frac{R_0^2+L_2^2}{2L_2}\right)}{\frac{R_0^2+L_1^2}{2L_1}\frac{R_0^2+L_2^2}{2L_2}(L_2-L_1)^3}}x\right) \mathrm{d}x \right|^2$$

$$J_{n,N,N'} = \left| \int_0^\infty \begin{array}{l} \dfrac{h_2(x)+h_1(x)-\frac{(h_1(x)-h_2(x))}{2\pi n}\sin\left(\frac{h_2(x)+h_1(x)}{h_1(x)-h_2(x)}2\pi n\right)}{h_2(x)-h_1(x)} \times \\ ch^\lambda\left(\dfrac{x}{\gamma}\right)\left[C_1\cdot{}_2F_1\left(u,v,\dfrac{1}{2};1-ch^2\left(\dfrac{x}{\gamma}\right)\right)+\right. \\ \qquad \left. +C_2\left(1-ch^2\left(\dfrac{x}{\gamma}\right)\right)^{\frac{1}{2}}\cdot{}_2F_1\left(u+\dfrac{1}{2},v+\dfrac{1}{2},\dfrac{3}{2};1-ch^2\left(\dfrac{x}{\gamma}\right)\right)\right]\times \\ ch^\lambda\left(\dfrac{x}{\gamma}\right)\left[C_3\cdot{}_2F_1\left(u,v,\dfrac{1}{2};1-ch^2\left(\dfrac{x}{\gamma}\right)\right)+\right. \\ \qquad \left. +C_4\left(1-ch^2\left(\dfrac{x}{\gamma}\right)\right)^{\frac{1}{2}}\cdot{}_2F_1\left(u+\dfrac{1}{2},v+\dfrac{1}{2},\dfrac{3}{2};1-ch^2\left(\dfrac{x}{\gamma}\right)\right)\right]\mathrm{d}x \end{array} \right|^2$$

## Appendix 2

Estimation of Relative Energy Error at Adiabatic Approximation

Let us define relative error for one-dimensional energy as ratio $\left(V_{\mathrm{Real}}(x)-\varepsilon_1^{\mathrm{Par(PT)}}(x)\right)/V_{\mathrm{Real}}(x)$, where $V_{\mathrm{Real}}(x)$ is exact calculated energy of CC in one-dimensional quantum well, $\varepsilon_1^{\mathrm{Par}}(x)\approx\alpha_1+\beta_1^2x^2$ is interpolated Taylor series of adiabatic approximated energy of CC and $\varepsilon_1^{\mathrm{PT}}(x)=\alpha_1-\frac{\lambda(\lambda-1)}{\gamma^2(ch(x/\gamma))^2}+\frac{\lambda(\lambda-1)}{\gamma^2}$ is the Pöschl-Teller approximated energy of CC, respectively. This estimation approach can be used as a useful tool for designing objects for practical applications from theoretically modeled samples. According to Fig. 9, at utilization of the adiabatic approximation the magnitude of the error comprises $10^{-4}$ what demonstrates the high accuracy one attains at implementation of this approximation.

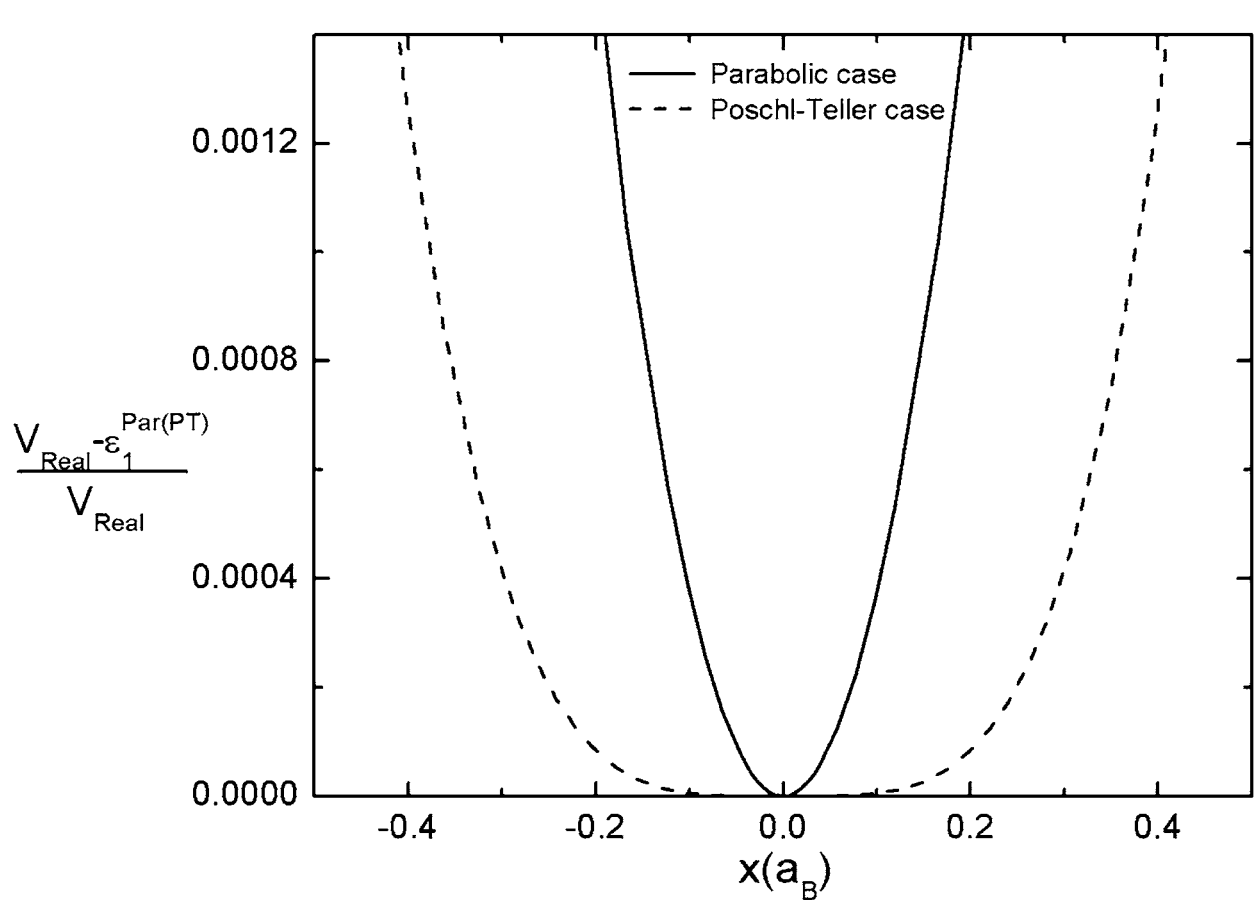

**Fig. 9** Estimation of relative energy error curve for both parabolic and Pöschl-Teller cases

## References

1. S. Fafard et al., Phys. Rev. B **50**, 8086 (1994). doi:10.1103/PhysRevB.50.8086
2. M. Grundmann et al., Phys. Rev. Lett. **74**, 4043 (1995). doi:10.1103/PhysRevLett.74.4043
3. S. Lee et al., Phys. Rev. B **61**, R2405 (2000). doi:10.1103/PhysRevB.61.R2405
4. A. Wojs et al., Phys. Rev. B **54**, 5604 (2001). doi:10.1103/PhysRevB.54.5604
5. A. Lorke et al., Phys. Rev. Lett. **84**, 2223 (2000). doi:10.1103/PhysRevLett.84.2223
6. L.C. Lew Yan Voon et al., J. Phys. Condens. Matter. **14**, 13667 (2002). doi:10.1088/0953-8984/14/49/321
7. K. Dvoyan, in *Proceedings of Semiconductor Micro- and Nanoelectronics the Fifth National Conference Aghveran*, Armenia, September 16–18 (2005), pp. 169–172
8. C. Trallero-Herrero et al., Phys. Rev. E. Stat. Nonlin. Soft Matter Phys. **64**, 056237–056241 (2001). doi:10.1103/PhysRevE.64.056237
9. A. Chanchapanyan, in *Proceedings of the Fifth International Conference "Semiconductor Micro- and Nanoelectronics"* (2005), p. 189
10. D. Leonard et al., Phys. Rev. B **50**, 11687 (1994). doi:10.1103/PhysRevB.50.11687
11. P. Harrison, *Quantum Wells, Wires and Dots: Theoretical and Computational Physics* (Wiley Ltd, NY, 2005)
12. S. Flügge, *Practical Quantum Mecannics* (Springer, Berlin, 1971)
13. Al.L. Efros, A.L. Efros, Sov. Phys. Semicond. **16**, 772 (1982)
14. H.A. Sarkisyan, Mod. Phys. Lett. **18**, 443 (2004). doi:10.1142/S0217984904007062